\documentclass[12pt]{article}
\usepackage{amsfonts}
\usepackage{amssymb}
\usepackage{graphicx}
\newcounter{rown}

\def\siz{\small}

\begin{document}
\title{ACCELERATION-EXTENDED GALILEAN SYMMETRIES WITH
CENTRAL CHARGES AND THEIR DYNAMICAL REALIZATIONS\footnote{Supported by the KBN grant 1P03B01828}}
\author{J. Lukierski$^{1)}$, P.C. Stichel$^{2)}$  and W.J.
Zakrzewski$^{3)}$
\\
\siz $^{1)}$Institute for Theoretical Physics,  University of
Wroc{\l}aw, \\ \siz pl. Maxa Borna 9,
 50--205 Wroc{\l}aw, Poland\\
 \siz e-mail: lukier@ift.uni.wroc.pl\\
\\ \siz
$^{2)}$An der Krebskuhle 21, D-33619 Bielefeld, Germany \\ \siz
e-mail:peter@physik.uni-bielefeld.de
\\ \\ \siz
$^{3)}$Department of Mathematical Sciences, University of Durham, \\
\siz Durham DH1 3LE, UK \\ \siz
 e-mail: W.J.Zakrzewski@durham.ac.uk
 }

\date{}
\maketitle

\begin{abstract}

We add to Galilean symmetries the transformations describing
constant accelerations. The corresponding extended Galilean algebra
allows, in any dimension $D=d+1$,  the introduction of one central
charge $c$ while in $D=2+1$ we can have three such charges: $c,
\theta$ and~$\theta'$. We present nonrelativistic classical
mechanics models, with higher order time derivatives and show that
they give dynamical realizations of our algebras. The presence of
central charge $c$ requires the acceleration square Lagrangian term.
We show that the general Lagrangian with three central charges can
be reinterpreted as describing an exotic planar particle coupled to
a dynamical electric and a constant magnetic field.
\end{abstract}

\section{Introduction}

In our recent paper \cite{rone} we studied conformal extensions of Galilean symmetries
in $D=2+1$ dimensions. Our extension was performed by the addition of four generators:

\noindent i) two generators - dilation $D$ and expansion $K$ - extending the Galilean algebra to the
Schr\"odinger algebra [2-4]

\noindent ii) two new generators $F_i$, ($i=1,2)$ describing constant accelerations.

It is easy to see that such an extension can be made for any nonrelativistic dimension $d$;
in fact, it can be obtained as a nonrelativistic $c\rightarrow \infty$ limit of the
relativistic $D$-dimensional ($D=d+1$) conformal algebra described by the Lie
algebra $O(d,2)$\footnote{In \protect\cite{rone} %%%[1]
 such a limit was mentioned for $D=2+1$}.
In such a derivation the generators $F_i$ ($i=1...d)$ are provided by the nonrelativistic
limits of space components of special conformal generators and describe, in a nonrelativistic
theory, constant accelerations.  Thus, such acceleration-extended Galilean symmetries, in $d$
spacial dimensions, are described by $\frac{1}{2}d(d-1)+3d+1$ generators, namely
($i=1 \ldots d$)
\begin{itemize}
\item $J_{ij}=-J_{ji}$ (space rotations $\alpha_{ij}$)
\item $P_i$ (space translations $a_i$)
\item $K_i$ (constant velocity motions, or Galilean boosts $v_i$)
\item $F_i$ (constant acceleration motions, or Galilean accelerations $b_i$)
\item $H$ (time translations $t'=t+a$).
\end{itemize}
These symmetries taken
 in their infinitesimal form
 can be realised in nonrelativistic space-time as
\begin{equation}\label{eone}
\delta x_i\,=\,-\delta a_i\,-\,\delta v_i t\,-\,\delta b_it^2\,+\,\delta \alpha_{ij}x_j,
\end{equation}
$$ \delta t\,=\,\delta a.$$

 The acceleration-extended Galilean algebra has the following nonvanishing commutators:
 $$[J_{ij},\,J_{kl}]\,=\,\delta_{ik}J_{jl}\,-\,\delta_{jl}J_{ik}\,+\,
 \quad (i\leftrightarrow j,k\leftrightarrow l),$$
 $$[J_{ij},\,A_{k}]\,=\,\delta_{ik}A_{j}\,-\,\delta_{jk}A_{i}
 \qquad \quad (A_i=P_i,K_i,F_i),$$
\begin{equation}\label{etwo}
[H,\,K_i]\,=\,P_i,
\end{equation}
\begin{equation}\nonumber
[H,\,F_i]\,=\,2K_i.
\end{equation}

The aim of this paper is to study central extensions of the algebra
 (\ref{etwo}-3) and its
dynamical representations.

 First we recall that the standard Galilean algebra
 (relations (\ref{etwo})
for $J_{ij},P_i,K_i$ and $H$) has two central extensions:

\noindent i) For arbitrary $D$ one can introduce one central extension \cite{rfive}
\begin{equation}\label{ethree}
[P_i,\,K_j]\,=\,m\delta_{ij},
\end{equation}
where $m$ describes the nonrelativistic mass parameter.

\noindent ii) In $D=2+1$ we can introduce a second charge $\theta$,
 called exotic \cite{rsix},\cite{rseven}
\begin{equation}\label{efour}
[K_i,\,K_j]\,=\,\theta \epsilon_{ij},
\end{equation}
which can be related to the noncommutativity of the space components of the
 nonrelativistic $D=2+1$ space-time \cite{reight}.

A dynamical realisation of the $D=2+1$ exotic algebra can be deduced
from the following nonrelativistic model \cite{reight}
\begin{equation} \label{efive}
L\,=\,\frac{1}{2} m\dot x_i^2\,-\,\frac{\theta}{2}\epsilon_{ij}\dot{x_i}\ddot{x_j}.
\end{equation}

One can show by considering the Jacobi identity for the three
 generators
  $H, P_i, F_j$ and using (3), (4)
 ($ 0 \equiv [H, [P_i, F_j]] +
  [P_i[F_j, H]] +
  [F_j, [H, P_i]] = - 2[P_i, K_j] = - 2m\delta_{ij}$)
  that in acceleration - extended Galilei algebra we must
 put $m=0$.
 In such a case one arrives at the following  central extensions
\begin{itemize}
\item For arbitrary $D$
 one  can introduce  a single central extension
\begin{equation}\label{esix}
[K_i,\,F_j]\,=\,2c\delta_{ij}.
\end{equation}
\item In $D=(2+1)$, in addition to relations
 (\ref{efour}) and (\ref{esix}) one can
have a third central charge
\begin{equation} \label{eseven}
[F_i,\,F_j]\,=\, \theta' \epsilon_{ij}.
\end{equation}
The algebraic consistency requires also that
\begin{equation} \label{eeight}
[P_i,F_j]\,=\,-2\theta \epsilon_{ij}.
\end{equation}
\end{itemize}

The paper is organised as follows. In the next section we consider a dynamical
 model in any dimension $D$ with the Lagrangian being given by
the square of the accelerations of the particle and the Noether charges satisfying (\ref{esix}).
 In Section 3 we discuss those properties of the acceleration-extended $D= 2+1$
  Galilean algebra which are model independent, {\it ie} have a purely algebraic origin.
 We consider the enveloping algebra
$U(\hat g_{\theta})$ with one central charge $\theta$ (see (5) and (9)) and we show that the symmetry algebras with three central charges can be embedded in $U(\hat g_{\theta})$.
 We also present the Casimirs in the presence of central charges and
discuss the possible enlargements of the symmetry algebra.
 A particular case here is the extension of the Galilean conformal algebra by the addition
 of two further generators $D$ and $K$. We find that when all three central parameters
 ($\theta, \theta'$ and $c$) are nonzero, the commutators in the Galilean conformal
 algebra involving the generators $D$ and $K$ are deformed. In Section 4 we introduce
  $D=2+1$ Lagrangian models which realize this three-fold centrally extended
 symmetry algebra. In particular, in Section 4.3, we show that the general planar model
with three central charges can be reinterpreted as describing the motion of a planar
noncommutative particle interacting with dynamic electric and constant magnetic fields. Section 5 contains some concluding remarks.

.

\section{An acceleration square Lagrangian and a new central charge $c$ (arbitrary $D=d+1$) }

The nonrelativistic kinetic term which is quasi-invariant under the acceleration
extended Galilei symmetry (1) has the form
($\ddot x^2=\ddot x_i\ddot x_i; \, i=1,..d$)
\begin{equation}
\label{eten}
L\,=\,\frac{c\,\ddot x^2}{2}.
\end{equation}
Indeed, performing the transformation (1) one obtains
\begin{equation}
\label{eeleven}
\delta L\,=\,-2c\,\delta b_i\,\ddot x_i\,=\,\frac{d}{dt}(-2c\,\delta b_i\,\dot x_i).
\end{equation}
Introducing $y_i=\dot x_i$ as an independent coordinate the Lagrangian (\ref{eten}) can be
put into the following equivalent first order form
\begin{equation}
\label{etwelve}
L\,=\,p_i(\dot x_i- y_i)\,+\,q_i\dot y_i\,-\,\frac{1}{2c}q_i^2.
\end{equation}
Using the Faddeev-Jackiw procedure \cite{rtwelve}-\cite{rthirteen}
we obtain the following nonvanishing Poisson brackets
\begin{equation}
\label{ethirteen}
\{x_i,\,p_j\}\,=\,\delta_{ij},\quad \{y_i,\,q_j\}\,=\,\delta_{ij}.
\end{equation}
The acceleration-extended Galilean transformations of the additional variables ($p_i,y_i,q_i)$,
which are consistent with the field equations
\begin{equation}
\label{efourteen}
y_i\,=\,\dot x_i,\qquad \dot p_i\,=\,0,
\end{equation}
$$ \dot y_i\,=\,\frac{1}{c} q_i,\qquad \dot q_i\,+\,p_i\,=\,0$$
take the form
$$ \delta p_i\,=\,\delta \alpha_{ij}\,p_j$$
\begin{equation}
\delta y_i\,=\,-\delta v_i\,-\,2t\,\delta b_i\,-\,\frac{\delta a}{c}q_i\,+\,\delta \alpha_{ij}\,y_j,
\label{efifteena}
\end{equation}
$$\delta q_i\,=\,-2c\,\delta b_i\,+\,\delta a\,p_i\,+\,\delta \alpha_{ij}\,q_j$$
and also
\begin{equation}
\label{efifteenb}
\delta x_i\,=\,-\delta a_i\,-\,t\,\delta v_i\,-\,t^2\,\delta b_i\,+\,\delta \alpha_{ij}x_j\, -\,y_i\,\delta a.
\end{equation}

 We note that the variation (\ref{efifteena}-\ref{efifteenb}) implies that
the Lagrangian is quasi-invariant
\begin{equation}
\label{eseventeen}
\delta L\,=\,\frac{d}{dt}\left(\delta a\,p_iy_i\,-2c\,\delta b_i\,y_i\right).
\end{equation}

The transformations (\ref{efifteena}-\ref{efifteenb}) are generated by the Poisson
brackets (\ref{ethirteen}) in the standard
way\footnote{The variation of the phase space variables
$Y_k$, generated by $G_r$, is given by $\delta Y_i=\delta \alpha_r\{G_r,Y_i\}$
 ($\alpha_r$ - symmetry parameters).} by the following realization
of the generators of the acceleration extended algebra with one central charge $c$ (see (7)):
\begin{equation}
\label{eeighteena}
 P_i\,=\,p_i,\qquad H\,=\,p_iy_i\,+\,\frac{1}{2c}\,q_i^2,
\end{equation}
\begin{equation}
J_{ij}\,=\,\frac{1}{2}\left(x_{[i}p_{j]}\,+\,y_{[i}q_{j]}\right),
\label{eeighteenb}
\end{equation}
\begin{equation}
\label{eeighteenc} K_i\,=\,q_i\,+\,p_it,\qquad
F_i\,=\,-2cy_i\,+\,2tq_i\,+\,t^2p_i \, ,
\end{equation}
where the second term in the energy generator $H$ is consistent with
 Souriau general theorem on the barycentric  decomposition of
 nonrelativistic systems in phase space [11,12].
It is easy to check, using (13), that (7) is satisfied.

In  quantum theory the quantized Poisson brackets (\ref{ethirteen}) take the form
\begin{equation}
[\hat x_i,\,\hat p_j]\,=\,i\hbar\delta_{ij},\qquad [\hat y_i,\,\hat q_j]\,=\,i\hbar\delta_{ij}
\label{etwenty}
\end{equation}
and can be realised in the standard way on the wave functions $\Psi(x_i,y_j)$ via
the Schr\"odinger representation:
\begin{equation}
\label{etwentyone}
\hat x_i=x_i,\quad \hat p_i\,=\,\frac{\hbar}{i}\frac{\partial}{\partial x_i},\quad \hat y_i=y_i,
\quad \hat q_i\,=\,\frac{\hbar}{i}\frac{\partial}{\partial y_i}.
\end{equation}
The Schr\"odinger equation, in the first quantized theory, corresponding to the Lagrangian
(\ref{eten}) and the Hamiltonian (\ref{eeighteena}) takes the form
\begin{equation}
\label{etwentytwo}
\left(-\frac{\hbar^2}{2c}\frac{\partial}{\partial y_i}\frac{\partial}{\partial y_i}
\,+\,\frac{\hbar}{i}\,y_i\frac{\partial}{\partial x_i}\right)\Psi\,=\,E\Psi.
\end{equation}

\section{Properties and Enlargements of the $D=2+1$ acceleration-extended Galilean algebra}

The space of dimension $d=2$ is special due to the existence of an antisymmetric covariant
constant tensor $\epsilon_{ij}=\left(\begin{array}{cc}0&1\\-1&0\end{array}\right)$. If,
for $d=2$, we extend the Galilei algebra by adding to it two generators $F_i$, describing
constant acceleration motions, due to the existence of the tensor $\epsilon_{ij}$, we
can introduce, through relations (\ref{efour}) and (\ref{eseven}) two independent
central charges $\theta$ and $\theta'$. In the general case, for $d=2$,
we can introduce three central
charges: $\theta$, $\theta'$ and $c$.

\subsection{How do we go from one central charge to three}

Let us observe that the status of all three central charges is not the same.
To see this we introduce the universal enveloping algebra $U(\hat g_{\theta})$, where
$\hat g_{\theta}$ denotes the acceleration-extended Galilei algebra with only one central
charge $\theta$. Then we observe that the following modification of the relations
\begin{eqnarray}
\left[ K_i,\,F_j\right] \,=\,0,&\quad \rightarrow\quad
\left[ K_i',\,F_j'\right] \,=\,2c\delta_{ij}\label{ethirtyeight}\\
\left[ F_i,\,F_j\right] \,=\,0,&\quad \rightarrow\quad
 \left[ F_i',\,F_j'\right] \,=\,\theta'\delta_{ij} \nonumber
\end{eqnarray}
can be achieved by the linear change of basis inside
the enveloping algebra $U(\hat g_{\theta})$
\footnote{Remaining generators $P_i,H,J,$ stay unchanged.}
\begin{eqnarray}
K_i'\,=&\,K_i\,+\,\frac{c}{2\theta}\,\epsilon_{ij}\,P_j\nonumber\\
F_i'\,=&\,F_i\,+\,\frac{c}{\theta}\,\epsilon_{ij}K_j\,+\,
\frac{1}{4}\left(\frac{c^2}{\theta^2}\,-\,
\frac{\theta'}{\theta}\right)P_i.\label{ethirtynine}
\end{eqnarray}
It can be checked that under this transformation all Lie algebra relations of
 $\hat g_{\theta}$ besides
(\ref{ethirtyeight}) remain unchanged. Thus we see that
 the two acceleration extended Galilean
algebra with a one parameter central extension (by the parameter $\theta$) and the
algebra with three central
extension parameters ($\theta, c, \theta'$) have the same enveloping algebras.
We can recall here the well known analogy: in $D=2+1$ spacetime the exotic Galilean algebra
with two central charges ($m,\theta$) can be obtained through the transformation
\cite{rfourteen},\cite{reight}
\begin{equation}
K_i'\,=\,K_i\,-\,\frac{\theta}{2m}\epsilon_{ij}P_j
\label{efourty}
\end{equation}
where $K_i$ and $P_i$ belong to the standard Galilean algebra with one central charge $m$:
\begin{equation}
[K_i,\,K_j]\,=\,[P_i,\,P_j]\,=\,0,\qquad [K_i,\,P_j]\,=\,i\delta_{ij}\,m.
\label{efourtyone}
\end{equation}
Indeed, the generators ($K_i',P'_i=P_i)$ in $D=2+1$ lead to the appearance of an `exotic'
central charge
\begin{equation}
[K_i',\,K_j']\,=\,i\epsilon_{ij}\theta,\quad [P_i',\,P_j']\,=\,0,\quad
[K_i',\,P_j']\,=\,i\delta_{ij}m,
\label{efourtytwo}
\end{equation}
which was recently reinterpreted as generating noncommutativity of the $d=2$ space coordinates
\cite{reight},\cite{rfifteen}.

\subsection{Casimirs}

$\hat g_{\theta}$ has two Casimirs:
\begin{eqnarray}
C_H\,=&\,H\,-\,\frac{1}{\theta}\,\epsilon_{ij}K_iP_j
\label{ntwentynine}\\
 \hbox{and}\qquad C_J\,=&\,J\,-\,\frac{1}{2\theta}(F_iP_i\,-\,K_i^2).\nonumber
 \end{eqnarray}
The Casimirs for the case of three central charges $\theta$, $\theta'$ and $c$ can be easily
obtained from (\ref{ntwentynine}) by the transformation (\ref{ethirtynine}) with
the result
\begin{eqnarray}
C'_H\,=&\,H\,-\,\frac{1}{\theta}\,\epsilon_{ij}K'_iP_j\,+\,\frac{c}{2\theta^2}\,P_i^2
\label{nthirty}\\
 \hbox{and}\qquad C'_J \, = & \,
  J \,- \,
  \frac{1}{2\theta}
 (F'_i {}
  P_i \,-\,
K'_i{}^2)
 \,-\,\frac{c}{\theta}H\,-\,\frac{\theta'}{8\theta^2}P_i^2.\nonumber
 \end{eqnarray}

 \subsection{Enlargement by an $O(2,1)$ algebra}

 We may add to $\hat g_{\theta}$ two further generators
 \begin{equation}
 \label{nthirtyone}
 J_{\pm}\,=\,\frac{1}{4\theta}\left(K_{\pm}^2\,-\,F_{\pm}P_{\pm}\right),
 \end{equation}
 where $K_{\pm}=K_1\pm iK_2$ {\it etc}. For $C_J=0$ they form together with $J$
 an $O(2,1)$ algebra
 \begin{equation}
 \label{nthirtytwo}
 [J_3,\,J_{\pm}]\,=\,\mp iJ_{\pm},\qquad [J_+,\,J_-]\,=\,2iJ_3,\end{equation}
 where $J_3=\frac{J}{2}$.
 The remaining nonvanishing commutators of $J_{\pm}$ describe the $O(2,1)$
 covariance of any two-vector $A_i\in(P_i,K_i,F_i)$
 \begin{equation}
 \label{nthirtythree}
 [J_+,\,A_-]\,=\,-iA_+,\quad [J_-,\,A_+]\,=\,iA_-.
 \end{equation}
In the Lagrangian model discussed in \cite{rone} we had $C_J=0$ and so this model
provided a realization of this $O(2,1)$ algebra.

\subsection{Extension to $D=2+1$ conformal Galilean algebra and central charges}

  To obtain the Galilean conformal algebra \cite{rone}
  one has to add to the acceleration-extended Galilean algebra (\ref{etwo}-\ref{ethree})
  two further generators: dilatation $D$ and expansion $K$, which together with the
  Hamiltonian, form an $O(2,1)$ subalgebra
  \begin{equation}
  [D,\,H]\,=\,-2H,\qquad [D,\,K]\,=\,2K,\qquad [H,\,K]\,=\,D.
  \label{efourtythree}
  \end{equation}
  The generators $D$ and $K$ are scalars\footnote{We recall that
  in $d=2$ $J_{ij}=\epsilon_{ij}J$.}
  \begin{equation}
  [D,\,J]\,=\,[K,\,J]\,=\,0
  \label{efourtyfour}
  \end{equation}
  and, in addition, they satisfy
  $$
  [D,\,P_i]\,=\,-P_i,\qquad [D,\,K_i]\,=\,0,\qquad [D,\,F_i]\,=\,F_i,$$
\begin{equation}
   [K,\,P_i]\,=\,-2K_i,\qquad [K,\,K_i]\,=\,-F_i,\qquad [K,\,F_i]\,=\,0.\label{efourtyfive}
 \end{equation}
  It is easy to argue that the Galilean conformal algebra does not permit the central
  extensions with parameters $c$ and $\theta'$. Indeed, if we observe that from (\ref{efourtythree}-\ref{efourtyfive})
  we get the following mass dimensions of the generators
 $$
  [P_i]\,=\,M^1,\qquad [K_i]\,=\,M^0,\qquad [F_i]\,=\,M^{-1}$$
\begin{equation}
  [H]\,=\,M^1,\quad [D]\,=\,M^0,\quad [K]\,=\,M^{-1}\quad [J]\,=\,M^0\label{efourtysix}
  \end{equation}
  we obtain from (\ref{efour}), (\ref{esix}) and (\ref{eseven}) that
  \begin{equation}
  [\theta]\,=\,M^0,\quad [c]\,=\,M^{-1},\quad [\theta']\,=\,M^{-2}.
  \label{efourtyseven}
  \end{equation}
  We see from (\ref{efourtyseven}) that the constants $c$ and $\theta'$ break the scale
  and conformal invariance. In fact, if we supplement the transformation (\ref{ethirtynine}) by the
  relations $D'=D$ and $K'=K$ we get from the generators of the acceleration-extended Galilean algebra
  $\hat g_{\theta}$ with one nonvanishing central charge the following deformation
  of the conformal Galilean algebra\footnote{We only list the modified relations.}
  \begin{eqnarray}
  \left[ K',\,P_i'\right] \,=&\,-2K_i'\,+\,\frac{c}{\theta}\epsilon_{ij}P_j'\nonumber\\
  \left[ K',\,F_i'\right] \,=&\,-\frac{c}{\theta}\epsilon_{ij}F_j'\,-\frac{3}{2}\frac{c^2}{\theta^2}K_i'
\,+\,\frac{1}{\theta^2}\left(\frac{c^2}{\theta}-\frac{\theta'}{2}\right)\epsilon_{ij}P_j\nonumber\\
\left[ K',\,K_i'\right] \,=&\,-F_i'\,+\,\frac{1}{4\theta}\left(\frac{c^2}{\theta}-\theta'\right)P_i'\label{efourtyeight}\\
\left[ D',\,K_i'\right] \,=&-\frac{c}{2\theta}\epsilon_{ij}P_j'\nonumber\\
\left[ D',\,F_i'\right] \,=&F_i'\,-\frac{c}{\theta}\epsilon_{ij}K_j'\,-\,\frac{1}{\theta}
\left(\frac{c^2}{\theta}-\frac{\theta'}{2}\right)P_i'.\nonumber
\end{eqnarray}
The parameters $c$ and $\theta'$ cease to be central charges
and take the role of deformation parameters (see {\it eg} \cite{rfifteena}).
The deformed conformal Galilean algebra with the generators $(P_i',K_i',F_i',J',H',D')$
and the three parameters ($\theta$, $c$ and $\theta'$) can be treated as a particular choice
of basis in the enveloping conformal Galilean algebra $U(\hat g_{\theta})$ with one central charge.

 \section{Dynamical planar models realizing the acceleration extended Galilean symmetry with three
 central charges}

  In this section we introduce and discuss $D=2+1$ nonrelativistic bilinear Lagrangian models which possess
  the acceleration-extended Galilean symmetry with three central charges ($\theta$, $\theta'$
  and $c$).
  \subsection{Higher-order derivative model}

The most general $D=2+1$ Lagrangian which is bilinear in the time derivatives
of the coordinates $x_i$ and is quasi-invariant
under transformations (\ref{eone}) has the form:
\begin{equation}
L\,=\,-\frac{\theta}{2}\,\epsilon_{ij}\,\dot x_i\,\ddot x_j\,+\,\frac{c}{2}\,\ddot x_j^2\,-\,
\frac{\theta'}{8}\,\epsilon_{ij}\ddot x_i\,\dot{\ddot{x_j}}.
\label{etwentysix}
\end{equation}
The model consisting only of the first term in (\ref{etwentysix}) ({\it i.e.} with $\theta\ne0$,
$\theta'=c=0$) was considered in detail in \cite{rone}.

 The extended Galilean transformations
(\ref{eone}) provide the following variation of the Lagrangian (\ref{etwentysix}):
\begin{equation}
\delta L\,=\,\frac{d}{dt}\left(-\frac{\theta}{2}\epsilon_{ij}\dot x_j\right)\,\delta v_i\,+\,\frac{d}{dt}\left[\theta\epsilon_{ij}\left(t\dot x_j\,-2x_j\right)\,-\,2c\dot x_i\,+\,\frac{\theta'}{4}
\epsilon_{ij}\ddot x_j\right]\,\delta b_i.
\label{etwentyseven}
\end{equation}
The invariance only modulo total derivative (quasi-invariance) implies the existence of central charges defined by the relations (\ref{efour}) and (\ref{esix}-\ref{eeight}).

In the first order formalism the Lagrangian (\ref{etwentysix}) then takes the form
\begin{equation}
L\,=\,p_i(\dot x_i-y_i)\,+\,q_i(\dot y_i-u_i)\,-\,\frac{\theta'}{8}\epsilon_{ij}\,u_i\,\dot u_j\,-\,
\frac{\theta}{2}\epsilon_{ij}y_i\dot y_j\,+\,\frac{c}{2}u^2.
\label{etwentyeight}
\end{equation}

The Euler-Lagrange equations (EOM) which follow from the Lagrangian (\ref{etwentyeight}) are
\begin{eqnarray}
y_i-\dot x_i\,=\,0, &\quad &\dot p_i\,=\, 0,\quad  u_i-\dot y_i\,=\,0, \label{ethritytwo}\\
p_i+\dot q_i\,+\theta \epsilon_{ij}\dot y_j\,=\,0,&\quad & q_i+\frac{\theta'}{4}\,\epsilon_{ij}\dot u_j\,-\,cu_i\,=\,0.
\nonumber\end{eqnarray}

Using the Faddeev-Jackiw procedure
  \cite{rtwelve}--\cite{rthirteen}
 we obtain  the following  nonvanishing
Poisson brackets:
\begin{eqnarray}
\{x_i,\,p_j\}\,&=\,\delta_{ij},\qquad \,\{y_i,\,q_j\}\,=\,&\delta_{ij}, \label{ethirtyone}\\
\{q_i,\,q_j\}\,&=\,-\theta\epsilon_{ij},\qquad \,\{u_i,\,u_j\}\,=\,&4\frac{\epsilon_{ij}}{\theta'}.
\nonumber\end{eqnarray}

To calculate from the Lagrangian (\ref{etwentyeight}) the Noether charges we  consider the following acceleration extended Galilean transformations  for a fixed
but arbitrary value of the time parameter $t$,
\begin{eqnarray} \delta x_i\,=&\,-\delta a_i\, -\, \delta v_it \,-\, \delta b_it^2\,+\,\delta \alpha\,\epsilon_{ij}\,x_j,\nonumber\\
\delta y_i\,=&\,-\delta v_i\, -\, 2t\delta b_i \,+\delta \alpha\,\epsilon_{ij}\,x_j,\nonumber\\
\delta u_i\,=&\,-2\delta b_i\, +\,\delta \alpha\,\epsilon_{ij}\,x_j,\label{ethirtyfive}\\
\delta p_i\,=&\,2\theta \epsilon_{ij}\delta b_j\, +\,\delta \alpha\,\epsilon_{ij}\,p_j,\nonumber\\
\delta q_i\,=&\,-2c \delta b_i\, +\,\delta \alpha\,\epsilon_{ij}q_j.\nonumber
\end{eqnarray}

These transformations leave the EOM (\ref{ethritytwo}) invariant.
 The generators which provide the transformations (\ref{ethirtyfive}) are
 obtained by the Noether theorem and are given by
\begin{eqnarray}
P_i\,=&\,p_i,\phantom{aaaaaaaaaaaaaaaaa}\nonumber\\
K_i\,=&p_it\,+\,q_i\,+\,\theta\,\epsilon_{ij}\,y_j,\phantom{aaaaaaaaaaaa}\nonumber\\
F_i\,=&-p_it^2\,+\,2K_it\,-\,2\theta \epsilon_{ij}x_j\,-\,2cy_i\,+\,\frac{\theta'}{2}\epsilon_{ij}u_j,\label{ethirtysix}\\
J\,=&\, \epsilon_{ij}x_ip_j\,+\,\epsilon_{ij}y_iq_j\,-\,\frac{\theta}{2}y^2\,-\,\frac{\theta'}{8}u^2\phantom{aaaaa}.
\nonumber
\end{eqnarray}

\subsection{Transmutation into a planar model with electromagnetic interaction}

Note that if we consider the velocities $y_i$ as the particle coordinates $X_i$ then the model
introduced in Section~4.1 would describe the interaction of noncommutative planar exotic particles
with dynamical electric fields and a constant magnetic field
(see \cite{rextra}, \cite{rnine}).

Let us perform the substitution:
\begin{eqnarray}
&p_i\to -E_i,\quad x_i\to \pi_i,\quad y_i\to X_i,\quad q_i\to P_i, \quad u_i\to Y_i,\nonumber\\
&\theta\to -B,\quad c\to m,\quad \hbox{and}\quad \theta'\to -4\theta m^2.
\label{nseventyfour}
\end{eqnarray}

Applying the substitution (\ref{nseventyfour}) into
the Lagrangian (42) we obtain

\begin{equation}
\label{nseventysix}
L\,=\,E_iX_i\,+\,\dot E_i\pi_i\,+\,P_i(\dot X_i-Y_i)\,+\,\frac{\theta m^2}{2}\epsilon_{ij}Y_i\dot Y_j
\,+\,\frac{B}{2}\epsilon_{ij}X_i\dot X_j\,+\,\frac{m}{2}Y^2.
\end{equation}

This is the Lagrangian of the higher-order derivative model
introduced in \cite{reight} supplemented by the standard
electromagnetic interaction with dynamical electric field $E_i$ and
a constant magnetic field $B$ and  a kinetic term $\dot E_i\pi_i$
added for the additional phase space variables $E_i$ and $\pi_i$. In
\cite{rextra}, \cite{rnine}  it was shown that such
electromagnetic couplings of planar particle models lead to the
electromagnetic enlargement of the Galilei symmetry with additional
generators given by the electric field $E_i$ and with three central
charges ($m,\theta$ and $B$). It can be checked that the Lagrangian
(\ref{nseventysix}) is quasi-invariant under the electromagnetically
enlarged $D=(2+1)$ Galilei group with the symmetry algebra which is
isomorphic to the acceleration-extended Galilei algebra with three
central charges $(\theta,\theta',c$)\footnote{In formulae (56-57) we
assume that the eletromagnetic coupling constant $e=1$.}.

Concluding we see that our massless planar
 higher order Lagrangian model (40)
  is clasically equivalent to the Lagrangian for
   a massive exotic planar point particle interacting
    with dynamical electric and constant magnetic fields.

\section{Final Remarks}

\begin{itemize}

\item We would like to add that the extension of the equivalent coordinate frames to the
ones with various  constant
 accelerations may  have possibly a physical
application based on recent developments in astrophysics.
 If in accordance with recent measurements (see {\it e.g.} \cite{rnine}-\cite{releven})
   the universe expands with increasing
velocity and approximately constant acceleration
  the coordinate frames extended by constant acceleration motions
may be useful in its description
\item Note that the Lagrangians invariant under acceleration-extended Galilean symmetries contain
higher derivatives. Although such Lagrangians have been discussed in literature
(see {\it eg} \cite{reighteen}, \cite{rnineteen}) their physical meaning is still unclear.
We would like, however, to point out that following our discussion in Section~4.2 these models might
have a meaning in the framework
of classical mechanics for exotic particles in $D=(2+1)$ dimensions.
\item

Analogously to the model introduced by us in \cite{reight} (cp. also \cite{rsixteen})
one can introduce for our model (\ref{etwentysix})  the phase space decomposition
into ``external'' and ``internal'' dynamics. This decomposition can be
 understood as a generalization of the Galilean-invariant decomposition of
 phase space into the center of mass and relative motion in standard
 classical mechanics. We plan to treat this method in a general
 framework and discuss the solutions of the present model and the
 interpretation of ``internal`` dynamics for the model with
 higher time derivatives  in a forthcoming publication.

\end{itemize}

\end{document}